\documentclass[fleqn,usenatbib,useAMS]{mnras}

\usepackage{graphicx}
\usepackage{xcolor}
\usepackage{color}
\usepackage{amsmath}
\usepackage{amssymb}
\usepackage{txfonts}
\usepackage{hyperref}
\usepackage{multirow}
\usepackage{notes2bib}
\usepackage{stfloats}
\usepackage{float}
\restylefloat{table}
\usepackage{caption} 
\usepackage{rotating}
\usepackage{pdflscape}
\usepackage{array}
\usepackage{hhline,booktabs,amsmath}
\usepackage{makecell}

\usepackage{newtxtext,newtxmath}

\hypersetup{colorlinks=true,linkcolor=[rgb]{1.,0.2,0.2},citecolor=[rgb]{0.1,0.4,1.},filecolor=[rgb]{0.7,0.2,0.2},urlcolor=[rgb]{0.0,0.2,1.}}
\definecolor{darkgreen}{rgb}{0.09, 0.45, 0.27}
\definecolor{amber(sae/ece)}{rgb}{1.0, 0.49, 0.0}

\defcitealias{Tortora+18_UCMGs}{T18}
\defcitealias{Scognamiglio20}{S20}
\defcitealias{Spiniello+24}{S24}
\defcitealias{Yan06}{Y06}
\defcitealias{MartinezParedes23}{MP23}
\def\INSPIRE{\mbox{{\tt INSPIRE}}}

\newcommand{\Reff}{$\mathrm{R}_{\mathrm{e}\,}$}
\newcommand{\Mstar}{M$_{\star}\,$}

\newcommand{\kms}{km s$^{-1}$}
\newcommand{\Msun}{M$_{\odot}\,$}

\usepackage[normalem]{ulem}

\usepackage[T1]{fontenc}
\usepackage{ae,aecompl}
\defcitealias{Spiniello+24}{S24}

\title[INSPIRE VIII. Emission lines and UV colours in ultra-compact massive galaxies]{ \centering INSPIRE: INvestigating Stellar Population In RElics VIII.\\ Emission lines and UV colours in ultra-compact massive galaxies}

\author[C. Spiniello]{Chiara~Spiniello$^{1}$\thanks{Contact e-mail: \href{mailto:chiara.spiniello@physics.ox.ac.uk}{chiara.spiniello@physics.ox.ac.uk}}, Mario~Radovich$^{2}$, Anna~Ferr\'e-Mateu$^{3,4}$, Roberto~De~Propris$^{5,6}$, Magda~Arnaboldi$^{7}$, \and
Francesco La Barbera$^{8}$,
Johanna~Hartke$^{5,9}$,
Giuseppe~D'Ago$^{10}$,
Crescenzo~Tortora$^{8}$, 
 Davide Bevacqua$^{11,12}$,  \and Michalina~Maksymowicz-Maciata$^{13}$, John~Mills$^{1}$, Nicola R. Napolitano$^{14,8}$, Claudia Pulsoni$^{15}$, \and Paolo Saracco$^{11}$, Diana Scognamiglio$^{16}$ %
\\
$^{1}$Sub-Dep. of Astrophysics, Dep. of Physics, University of Oxford, Denys Wilkinson Building, Keble Road, Oxford OX1 3RH, United Kingdom\\
$^{2}$INAF - Osservatorio astronomico di Padova, Vicolo Osservatorio 5, I-35122 Padova, Italy\\
$^{3}$ Instituto de Astrof\'isica de Canarias, V\'ia L\'actea s/n, E-38205 La Laguna, Tenerife, Spain\\
$^{4}$ Departamento de Astrof\'isica, Universidad de La Laguna, E-38200, La Laguna, Tenerife, Spain\\
$^{5}$Finnish Centre for Astronomy with ESO (FINCA), FI-20014 University of Turku, Finland\\
$^{6}$Department of Physics and Astronomy, Botswana International University of Science and Technology, Private Bag 16, Palapye, Botswana\\
$^{7}$European Southern Observatory,  Karl-Schwarzschild-Stra\ss{}e 2, 85748, Garching, Germany\\
$^{8}$INAF -  Osservatorio Astronomico di Capodimonte, Via Moiariello  16, 80131, Naples, Italy\\
$^{9}$Tuorla Observatory, Department of Physics and Astronomy, University of Turku, 20014 Turku, Finland\\
$^{10}$Institute of Astronomy, University of Cambridge, Madingley Road, Cambridge CB3 0HA, United Kingdom\\
$^{11}$INAF - Osservatorio Astronomico di Brera, via Brera 28, 20121 Milano, Italy\\
$^{12}$DiSAT, Universitá degli Studi dell’Insubria, via Valleggio 11, I-22100 Como, Italy\\
$^{13}$Astrophysics Group, H. H. Wills Physics Laboratory, University of Bristol, Tyndall Avenue, Bristol, BS1 8TL, United Kingdom\\
$^{14}$Department of Physics E. Pancini, University Federico II, Via Cinthia 21, I-80126, Naples, Italy\\
$^{15}$Max-Planck-Institut f\"{u}r  extraterrestrische Physik, Giessenbachstrasse, 85748 Garching, Germany\\
$^{16}$Jet Propulsion Laboratory, California Institute of Technology, 4800,  Oak Grove Drive - Pasadena, CA 91109, USA}

\date{Last updated 2025 March 26}

\pubyear{{\the\year{2025}}}

\begin{document}
\label{firstpage}
\pagerange{\pageref{firstpage}--\pageref{lastpage}}
\maketitle

\begin{abstract}
We report the discovery of emission lines in the optical spectra of ultra-compact massive galaxies (UCMGs) from \INSPIRE\, including relics, which are the oldest galaxies in the Universe. 
Emission-lines diagnostic diagrams suggest that all these UCMGs, independently of their star formation histories, are `retired galaxies'. They are inconsistent with being star-forming but lie in the same region of shock-driven emissions or
photoionisation models, incorporating the contribution from post-asymptotic giant branch (pAGB) stars. 
Furthermore, all but one \INSPIRE\ objects have a high [OII]/H$\alpha$ ratio, resembling what has been reported for normal-size red and dead galaxies. 
The remaining object (J1142+0012) is the only one to show clear evidence for strong active galactic nucleus activity from its spectrum. 
We also provide near-UV (far-UV) fluxes for 20 (5) \INSPIRE objects that match in GALEX. Their $NUV-r$ colours are consistent with those of galaxies lying in the UV green valley, but also with the presence of
recently ($\le0.5$ Gyr) formed stars at the sub-percent fraction level.
This central recent star formation could have been ignited by gas that was originally ejected during the pAGB phases and then re-compressed and brought to the core by the ram-pressure stripping of Planetary Nebula envelopes. Once in the centre, it can be shocked and re-emit spectral lines. 
\end{abstract}




\section{Introduction}
Relic galaxies \citep{Trujillo+09_superdense, Trujillo14, Ferre-Mateu+17} 
are the local descendants of high-redshift red nuggets that have completely missed the size-growth evolutionary phase \citep{Daddi+05, Trujillo+07, Buitrago+08,  vanDokkum+08, Naab+14} and have evolved passively and undisturbed from their first intense and fast high-$z$ star formation (SF) burst. 
Since they are made almost exclusively of “in situ” very old stars, like the innermost regions of massive galaxies (e.g., \citealt{LaBarbera+19, Barbosa21}) they provide a unique opportunity to track the evolution of this stellar component, which is mixed with the accreted one in normal early-type galaxies (ETGs). They are the ideal systems to investigate and understand the mass assembly in the early Universe with the amount of detail currently available only for galaxies in the local Universe. 
Moreover, since the number density of relics and its redshift evolution depends strongly on the processes acting during the size growth and how they are modeled, counting relics at low-z is an incredibly valuable way to disentangle between different galaxy evolution models. 

Recent claims have reported the presence of a subpercent fraction of young stellar populations in the innermost region of very massive galaxies and bright cluster galaxies \citep{Salvador-Rusinol2021_nature}. This is the region where the pristine (i.e., in situ), oldest stars should dominate the light budget \citep{Barbosa21}. Moreover, the same amount of younger stars ($\sim1\%$) have also been found in the most extreme relic in the local Universe, NGC~1277 \citep{Salvador-Rusinol22}, by fitting near-ultraviolet (NUV) and optical line-strength indices. Since NGC~1277 has not experienced any mergers or interactions with other galaxies, the presence of younger stars in its centre indicates that the gas for the formation of these new stars must be associated with the galaxy itself, i.e., with intrinsic processes rather than external ones. 
\citet{Dopita00} highlight that the
presence of gas and dust at the centres of passive galaxies may originate from the mass loss of evolved stars.  According to their simulation and analysis, in high-pressure interstellar environments like the centres of massive ETGs or UCMGs and relics, the material in the shocked shell of planetary nebulae (PNe) will cool and its expansion reversed, causing a collapse into the denser central region. This will, in turn, recompress the dusty material ejected during the AGB phase of stellar evolution and allow the cold dusty clouds to fall intact (without being destroyed by the hot interstellar medium) towards the nucleus of the galaxy. During this process, the gas could be shocked, creating shock-driven emission lines and, in some cases, form a small percentage of new stars, as in NGC~1277.  

The paper is organised as follows. In Section~\ref{sec:sample} we briefly describe the data used in this paper, highlighting the definition and selection of ultra-compact massive galaxies (UCMGs) and the confirmation of their relic nature. In Section~\ref{sec:emissionline} we analyse the emission lines, obtain ratios, and plot diagnostic diagrams to investigate their origins. In Section~\ref{sec:UV} we focus on the UV, analysing the $NUV-r$ colours. We then discuss the findings and conclude in Section~\ref{sec:discussion} investigating which could be the possible source of both colour and emission lines. 

Throughout the paper, we assume a standard $\Lambda$CDM cosmology with $H_0$=$69.6$ \kms Mpc$^{-1}$, $\Omega_{\mathrm{\Lambda}}$=$0.714$ and $\Omega_{\mathrm{M}}$=$ 0.286$ \citep{Bennett14}.


\section{The sample}
\label{sec:sample}
In this paper, we leverage the INvestigating Stellar Population In RElics (\INSPIRE, \citealt{Spiniello+21, DAgo23, Spiniello+24}) dataset. 
\INSPIRE\ is based on an ESO Large Programme that targets 52 spectroscopically confirmed UCMGs at $0.1<z<0.4$ with the X-Shooter spectrograph (XSH, \citealt{Vernet11}). 
These objects were initially found in \citet{Tortora+16_compacts_KiDS} from the Kilo Degree Survey (KiDS, \citealt{Kuijken11}) third data release (DR3, \citealt{deJong+17_KiDS_DR3}), and then spectroscopically confirmed in \citealt{Tortora+18_UCMGs} and \citealt{Scognamiglio20} (hereafter T18 and S20, respectively) with low signal-to-noise (SNR) and medium-resolution optical spectra. 
For each galaxy, structural parameters and stellar masses have been derived from $ugri$ KiDS photometry \citep{Roy+18, Tortora+18_UCMGs, Scognamiglio20}. 
All objects are UCMGs in the sense that they are clear outliers in the stellar mass-size plane (see Fig.~2 in \citealt{Spiniello20_Pilot}), having very small sizes (with effective radii \Reff$<2$ kpc) with respect to the overall population of passive galaxies with relatively large stellar masses (\Mstar$>6\times10^{10}$\Msun). 
Stellar masses and sizes for the entire \INSPIRE\ sample are provided in Table~1 of \citet[][hereafter S24]{Spiniello+24}.

From UVB-to-NIR high-resolution, medium signal-to-noise (SNR) spectra\footnote{Publicly available through the \href{https://archive.eso.org/scienceportal/home?data_collection=INSPIRE}{ESO Phase 3 Science Archive}} we have measured the integrated stellar kinematics \citep{DAgo23}, as well as stellar population age, metallicity, [Mg/Fe] abundances and the IMF slope \citep{Martin-Navarro+23, Maksymowicz-Maciata24}.  Of these UCMGs, 38 have been classified as relics, since they formed more than 75\% of their stellar masses already by $z>2$ \citep{Spiniello+21, Spiniello+24}. Hence, \INSPIRE\ 
has enlarged by at least a factor of $5$ the number of nearby fully confirmed relics, providing the first statistically significant sample of these objects. 
Here, we focus on optical emission lines and UV colours, aiming at investigating on their sources and on whether a correlation exists between UV-light, emission lines and the \textit{degree of relicness} (DoR). 
The DoR is a dimensionless number ranging from 0 (non relic) to 1 (extreme relic), that quantifies how extreme and peaked the SFH of these UCMGs is. We refer the reader to \citet[ hereafter S24]{Spiniello+24} for more details on how the DoR is computed. 

We use here the 52 INSPIRE UVB+VIS XSH spectra to measure emission lines, as described in the next section. 

\section{Emission lines}
\label{sec:emissionline}
In roughly half of the spectra released in the third \INSPIRE\ data release (DR3, \citetalias{Spiniello+24}), we observe convincing evidence for emission lines from [OII] ($\lambda\sim3727$\AA) and [NII] ($\lambda\sim6583$\AA) in the combined UVB+VIS spectra.   
This is broadly consistent with the statistics, based on a much larger sample of normal-sized galaxies ($\sim300,000$ galaxies from SDSS DR4, \citealt{Adelman06}), of \citet{Yan06}: 38\% of all red galaxies have detectable [OII] emission. 

Weak emissions are also detected in [OIII] ($\lambda\sim5007$\AA) and [SII] ($\lambda\sim6720$\AA).  
Finally, H$\beta$ and H$\alpha$ also have very weak emission components in some cases, although they are contaminated by stellar absorption lines. No other emission lines are detected in the wavelength range [2700-9500]\AA.

Unfortunately, the SNRs of most emission lines are generally low ($\le10$), with the exception of J1142+0012 (which will be discussed later). Moreover, the continuum underlying [OII] has many narrower absorption features that could bias the measurement of the flux. 
Hence, to compute the equivalent width (EW) of the line and flux in emission, we first subtract the stellar component and then perform a fit to the emission lines. 

To compute EWs and fluxes, we use a Python code, based on the \textsc{LMFIT} library\footnote{\url{https://lmfit.github.io/lmfit-py/}}, that fits emission lines with Gaussian profiles. We also note that for [OII]($\lambda$3736,29), H$\alpha$+[NII]($\lambda$6548,83) and [SII]($\lambda$6716,31), the fit is performed using several Gaussians simultaneously.  
However, in order to ensure that the results are based only on reliable measurements, we
consider only lines with an SNR$>2$ and a FWHM$>1$\AA. 

The fluxes of all emission lines that pass the above thresholds are provided in the central block of columns in Table~\ref{tab:emission}, along with their uncertainties, in units of $10^{-17}$ erg s$^{-1}$cm$^{-2}$\AA$^{-1}$. 
We note that the velocity dispersion of the emission lines is similar to that of the stars, which disfavour an origin due to AGN or extreme shock waves. 

In the following, we will make use of line-line plots and emission line diagnostic to investigate the possible origin of emission in UCMGs and relics. 
\begin{figure}
\includegraphics[width=\columnwidth]{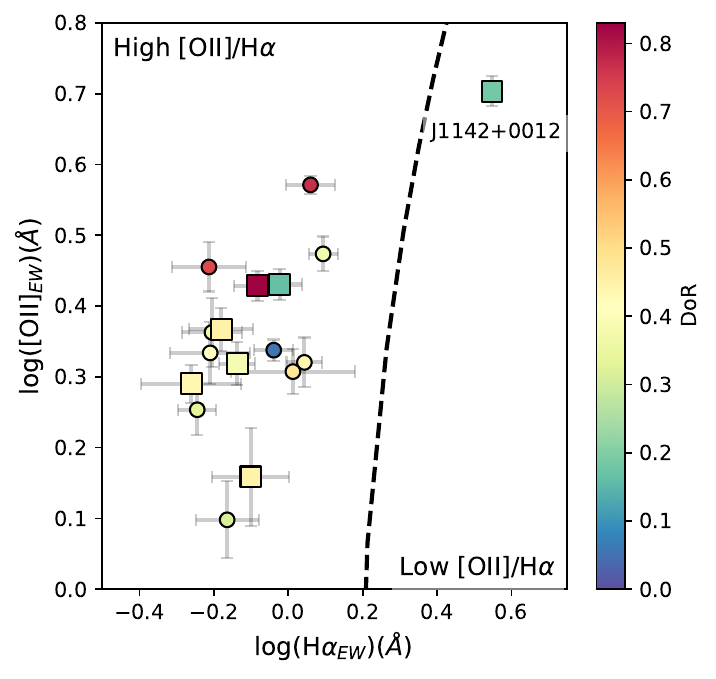}
\caption{The distribution of \INSPIRE\ galaxies in $\log$H$\alpha$ -- $\log$[OII] EWs, and colour-coded by DoR. Objects on the left side of the black dashed lines are classified as 'High [OII]/H$\alpha$'.  
The only 'Low [OII]/H$\alpha$' galaxy from the \INSPIRE\ sample is J1142+0122, which shows clear sign of AGN activity in its spectrum. Squares denote objects with a match in GALEX.}
\label{fig:yan}
\end{figure}

\begin{table*}
\small
\centering
\begin{tabular}{cc|cccccc|ccc}
\hline  
  \multicolumn{1}{c}{GALAXY} &
  \multicolumn{1}{c|}{DoR} &
  \multicolumn{1}{c}{[OII]} &
  \multicolumn{1}{c}{H$\beta$} &
  \multicolumn{1}{c}{[OIII]} &
  \multicolumn{1}{c}{H$\alpha$} &
  \multicolumn{1}{c}{[NII]} &
  \multicolumn{1}{c}{[SII]} &
  \multicolumn{1}{|c}{FUV} &
  \multicolumn{1}{c}{NUV} &
  \multicolumn{1}{c}{NUV-r} \\
  \multicolumn{1}{c}{ID} &
  \multicolumn{1}{c|}{ } &
  \multicolumn{1}{c}{$\lambda3727+3729$} &
  \multicolumn{1}{c}{$\lambda4861$} &
  \multicolumn{1}{c}{$\lambda5007$} &
  \multicolumn{1}{c}{$\lambda6563$} &
  \multicolumn{1}{c}{$\lambda6583$} &
  \multicolumn{1}{c}{$\lambda6717+6731$} &
  \multicolumn{1}{|c}{(mag)} &
  \multicolumn{1}{c}{(mag)} &
  \multicolumn{1}{c}{ } \\
\hline
J0211-3155 & 0.72 & $0.72\pm0.19$ & $0.57\pm0.21$ & $0.50\pm0.18$ &     --       &     --       &     --       &     --       & 23.42 & 3.64\\
J0224-3143 & 0.56 &     --       &     --       &     --       &     --       &     --       &     --       &     --       &     --       &     --      \\
J0226-3158 & 0.12 &     --       &     --       & $1.21\pm0.35$ &     --       &     --       &     --       &     --       & 23.27 & 4.02\\
J0240-3141 & 0.43 & $0.87\pm0.21$ & $0.72\pm0.42$ & $2.23\pm0.73$ &     --       &     --       &     --       &     --       &     --       &     --      \\
J0314-3215 & 0.42 & $2.03\pm0.21$ & $0.59\pm0.28$ & $1.47\pm0.46$ & $1.09\pm0.27$ & $1.90\pm0.31$ & $1.34\pm0.24$ &     --       &     --       &     --      \\
J0316-2953 & 0.40 & $1.36\pm0.19$ &     --       &     --       &     --       & $1.55\pm0.30$ &     --       &     --       & 22.87 & 3.21\\
J0317-2957 & 0.51 &     --       &     --       &     --       &     --       &     --       &     --       &     --       &     --       &     --      \\
J0321-3213 & 0.37 & $6.10\pm0.34$ & $1.03\pm0.32$ & $1.02\pm0.34$ & $3.40\pm0.30$ & $4.27\pm0.34$ &     --       &     --       &     --       &     --      \\
J0326-3303 & 0.25 &     --       & $0.84\pm0.28$ &     --       &     --       &     --       &     --       &     --       &     --       &     --      \\
J0838+0052 & 0.54 & $1.39\pm0.18$ &     --       &     --       &     --       & $1.18\pm0.30$ &     --       &     --       &     --       &     --      \\
J0842+0059 & 0.73 & $2.72\pm0.22$ & $2.10\pm0.53$ & $0.77\pm0.34$ & $1.24\pm0.28$ & $2.00\pm0.32$ &     --       &     --       &     --       &     --      \\
J0844+0148 & 0.45 & $1.32\pm0.21$ &     --       & $2.42\pm0.68$ & $1.26\pm0.30$ & $2.09\pm0.36$ &     --       &     --       & 22.26 & 2.48\\
J0847+0112 & 0.83 & $5.09\pm0.25$ & $0.94\pm0.69$ & $3.30\pm0.97$ & $3.18\pm0.46$ & $3.15\pm0.48$ & $4.72\pm0.71$ &     --       & 23.40 & 4.99\\
J0857-0108 & 0.39 &     --       &     --       & $0.83\pm0.21$ &     --       &     --       &     --       &     --       & 24.07 & 4.86\\
J0904-0018 & 0.32 &     --       &     --       &     --       &     --       &     --       &     --       & 23.17 & 22.20 & 3.09\\
J0909+0147 & 0.79 & $4.74\pm0.64$ & $2.55\pm0.76$ &     --       &     --       &     --       &     --       &     --       &     --       &     --      \\
J0917-0123 & 0.44 & $2.79\pm0.23$ & $0.69\pm0.32$ & $1.59\pm0.29$ & $3.17\pm0.35$ & $5.36\pm0.41$ & $4.43\pm0.54$ &     --       &     --       &     --      \\
J0918+0122 & 0.43 &     --       & $0.83\pm0.49$ &     --       &     --       & $1.86\pm0.46$ &     --       &     --       & 24.12 & 4.99\\
J0920+0126 & 0.25 &     --       &     --       & $0.85\pm0.29$ &     --       & $0.72\pm0.28$ &     --       &     --       &     --       &     --      \\
J0920+0212 & 0.64 &     --       &     --       & $3.03\pm0.97$ &     --       &     --       &     --       & 22.77 & 22.27 & 3.40\\
J1026+0033 & 0.29 & $4.17\pm0.50$ &     --       & $4.65\pm1.77$ &     --       & $4.11\pm1000.00$ &     --       & 22.92 & 21.58 & 4.19\\
J1040+0056 & 0.77 & $6.05\pm0.18$ &     --       & $4.44\pm0.38$ & $2.62\pm0.40$ & $2.76\pm0.42$ & $3.89\pm0.34$ &     --       &     --       &     --      \\
J1114+0039 & 0.40 &     --       &     --       &     --       &     --       &     --       &     --       &     --       &     --       &     --      \\
J1128-0153 & 0.34 & $1.47\pm0.32$ &     --       & $2.28\pm0.44$ &     --       &     --       &     --       &     --       & 23.29 & 4.73\\
J1142+0012 & 0.18 & $46.05\pm2.23$ & $21.16\pm0.76$ & $173.99\pm4.94$ & $93.04\pm4.27$ & $162.70\pm4.90$ & $88.35\pm3.59$ & 21.97 & 21.34 & 4.32\\
J1154-0016 & 0.11 &     --       &     --       &     --       &     --       &     --       &     --       &     --       & 23.71 & 4.19\\
J1156-0023 & 0.30 & $2.50\pm0.24$ & $1.18\pm0.58$ & $2.63\pm0.79$ &     --       & $2.10\pm0.57$ &     --       &     --       &     --       &     --      \\
J1202+0251 & 0.36 &     --       &     --       &     --       &     --       &     --       &     --       &     --       &     --       &     --      \\
J1218+0232 & 0.45 & $3.47\pm0.24$ &     --       & $2.07\pm1.94$ & $1.66\pm0.33$ & $2.68\pm0.38$ &     --       &     --       & 23.23 & 4.00\\
J1228-0153 & 0.39 & $2.07\pm0.29$ & $0.91\pm0.35$ & $0.78\pm0.36$ &     --       & $1.36\pm0.34$ &     --       &     --       &     --       &     --      \\
J1402+0117 & 0.31 & $1.08\pm0.14$ & $1.10\pm0.41$ & $1.16\pm0.42$ & $1.14\pm0.22$ & $1.33\pm0.24$ &     --       &     --       &     --       &     --      \\
J1411+0233 & 0.41 & $1.54\pm0.22$ &     --       &     --       &     --       &     --       &     --       &     --       &     --       &     --      \\
J1412-0020 & 0.61 &     --       &     --       &     --       &     --       &     --       &     --       &     --       & 22.54 & 3.35\\
J1414+0004 & 0.36 & $3.09\pm0.34$ & $1.20\pm0.91$ & $1.41\pm0.32$ & $1.71\pm0.32$ & $3.85\pm0.38$ &     --       &     --       &     --       &     --      \\
J1417+0106 & 0.33 &     --       &     --       & $0.65\pm0.38$ &     --       &     --       &     --       &     --       &     --       &     --      \\
J1420-0035 & 0.41 &     --       &     --       & $1.07\pm0.59$ &     --       &     --       & $2.39\pm0.55$ &     --       &     --       &     --      \\
J1436+0007 & 0.33 & $4.15\pm0.35$ &     --       &     --       & $2.75\pm0.33$ & $4.21\pm0.36$ & $2.43\pm0.46$ &     --       &     --       &     --      \\
J1438-0127 & 0.78 & $0.83\pm0.15$ & $0.65\pm0.28$ & $1.64\pm0.35$ &     --       & $1.22\pm0.28$ &     --       &     --       &     --       &     --      \\
J1447-0149 & 0.38 & $4.94\pm0.34$ & $1.99\pm0.65$ & $3.05\pm0.66$ & $3.21\pm0.35$ & $4.75\pm0.39$ & $3.45\pm0.48$ &     --       & 23.32 & 4.71\\
J1449-0138 & 0.60 &     --       & $0.65\pm0.33$ & $0.34\pm0.14$ & $0.52\pm0.16$ &     --       &     --       &     --       &     --       &     --      \\
J1456+0020 & 0.17 & $4.06\pm0.20$ & $1.06\pm0.60$ & $0.88\pm0.37$ & $2.10\pm0.29$ & $4.26\pm0.35$ &     --       & 23.02 & 23.18 & 3.72\\
J1457-0140 & 0.47 &     --       &     --       &     --       &     --       &     --       &     --       &     --       & 23.53 & 4.10\\
J1527-0012 & 0.38 &     --       &     --       &     --       &     --       &     --       &     --       &     --       &     --       &     --      \\
J1527-0023 & 0.37 & $2.26\pm0.21$ &     --       &     --       &     --       & $2.51\pm0.62$ &     --       &     --       & 23.99 & 4.35\\
J2202-3101 & 0.48 & $2.66\pm0.19$ & $2.30\pm0.79$ & $2.28\pm0.50$ & $2.62\pm1.00$ & $3.23\pm1.11$ &     --       &     --       &     --       &     --      \\
J2204-3112 & 0.78 &     --       & $1.05\pm0.37$ & $0.44\pm0.16$ &     --       &     --       &     --       &     --       &     --       &     --      \\
J2257-3306 & 0.27 &     --       &     --       & $0.59\pm0.24$ &     --       &     --       &     --       &     --       &     --       &     --      \\
J2305-3436 & 0.80 &     --       &     --       & $0.45\pm0.26$ &     --       &     --       &     --       &     --       &     --       &     --      \\
J2312-3438 & 0.36 &     --       &     --       & $0.53\pm0.18$ &     --       &     --       &     --       &     --       &     --       &     --      \\
J2327-3312 & 0.06 & $3.96\pm0.14$ &     --       & $2.57\pm0.25$ & $2.34\pm0.28$ & $4.65\pm0.34$ & $3.00\pm0.73$ &     --       &     --       &     --      \\
J2356-3332 & 0.44 & $2.41\pm0.15$ & $1.27\pm0.40$ & $1.94\pm0.39$ & $1.03\pm0.32$ & $1.07\pm0.32$ &     --       &     --       & 23.27 & 3.46\\
J2359-3320 & 0.71 &     --       &     --       & $2.79\pm1.00$ &     --       &     --       &     --       &     --       &     --       &     --      \\
\hline
\end{tabular}   
\caption{Emission lines fluxes for the \INSPIRE\ galaxies. 
For each object, we give ID and DoR (computed in \citetalias{Spiniello+24}) in the first block of columns, the flux with uncertainties of all the lines we use in this letter in the second block, and GALEX observed magnitudes and the $NUV-r$ colour in the third block. Missing measurements are not passing the thresholds described in 
Sec.~\ref{sec:emissionline}. Units for spectra are in $10^{-17}$ erg s$^{-1}$cm$^{-2}$\AA$^{-1}$.}
    \label{tab:emission}
\end{table*}

\subsection{[OII]/H$\alpha$ bimodality}
\citet[ hereafter Y06]{Yan06} reported the discovery of a bimodality in [OII]/ H$\alpha$ ratio among $\sim$300,000 galaxies from the fourth SDSS Data Release (DR4; \citealt{Adelman06}). One mode is largely associated with star-forming galaxies, while the other mode contains galaxies with line ratios compatible with low-ionisation nuclear emission-line regions (LINERs). Narrow-line Seyferts and transition objects mostly fall in between the two dominant populations, consistent with the picture that both SF and AGNs (or some other sources, e.g. pAGBs) might contribute substantially to the emission in these objects. 
In Figure~\ref{fig:yan}, we reproduce the $\log$([OII]$_{EW}$) - $\log$(H$\alpha_{EW}$) diagnostic plot used by \citetalias{Yan06} for the \INSPIRE\ galaxies\footnote{Note that, in Figure~\ref{fig:yan}, we plot the EWs of the lines, rather than their fluxes.} to check whether a bimodality exists also in this sample (i.e., with objects with higher DoR falling in the LINERs region while nonrelic UCMGs being more compatible with star-forming galaxies). 


All but one object are classified as 'High [OII]/H$\alpha$' galaxies, according to the threshold set by Eq.~6 in \citetalias{Yan06}. Indeed, \citetalias{Yan06} showed that the bimodality in [OII]/H$\alpha$ ratio echoes the galaxies' colour bimodality and the great majority of red galaxies reside in the 'High [OII]/H$\alpha$' category. Moreover, even though, in theory, SF models can explain a high [OII]/H$\alpha$ ratio, they would predict much lower [NII]/[OII] ratios (for solar or supersolar metallicities) than the one observed for the \INSPIRE\ sample \citep{Gutkin16}. 

The only system classified as 'Low [OII]/H$\alpha$' is J1142+0012 (DoR$=0.18$, i.e. nonrelic). 
This is the only \INSPIRE\ object that has a spectrum consistent with a  Seyfert~1 AGN, with broad and strong H$\alpha$ emission. This system will be further investigated in an upcoming publication. 
The three extreme relics for which a secure measurement of the [OII] and H$\alpha$ lines is possible seem to have a higher $\log$([OII]$_{EW}$) than objects with lower DoR. However, unfortunately, we do not have a large enough sample to draw firm conclusions.



\begin{figure*}
\includegraphics[width=\textwidth]{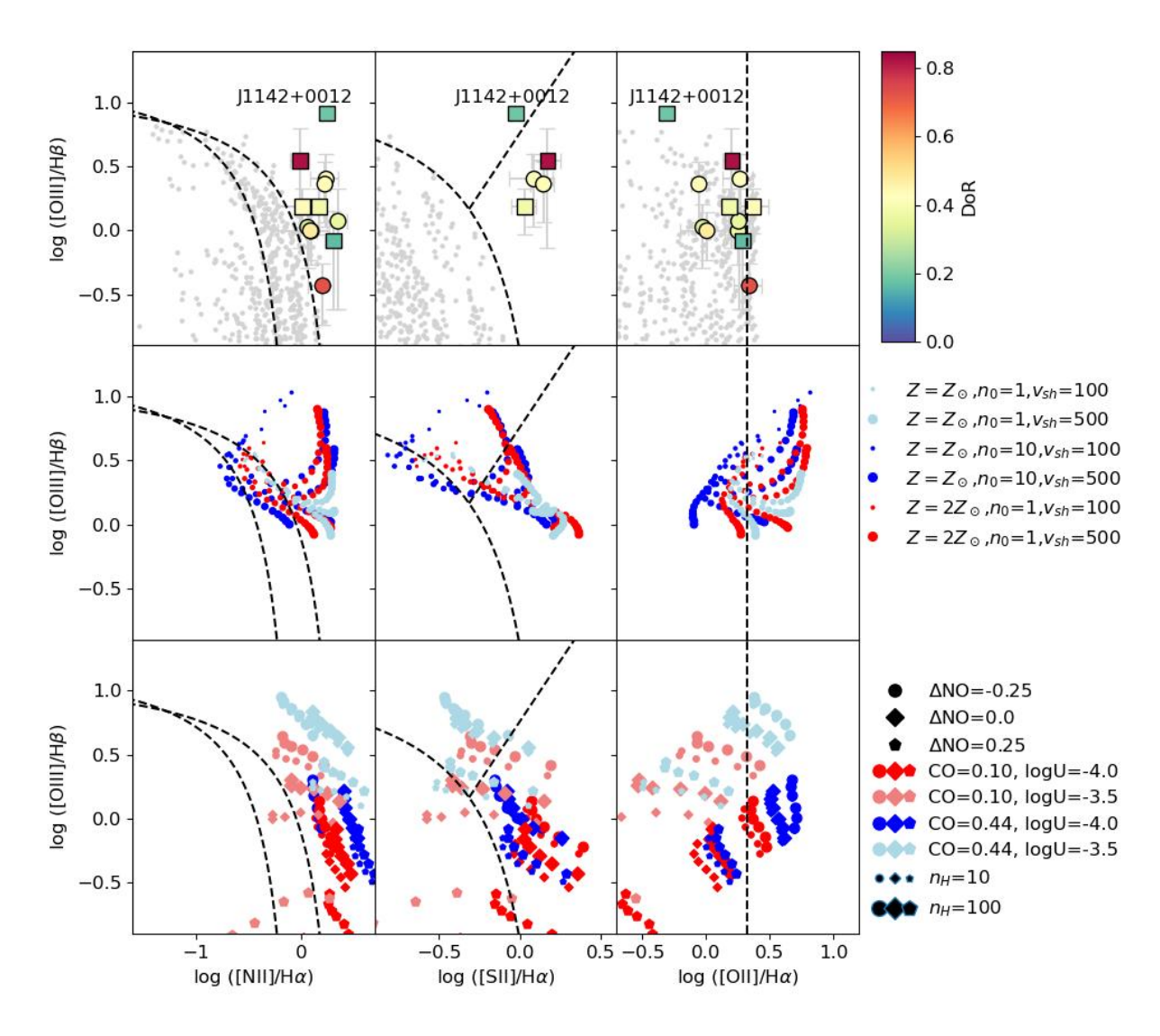}
\caption{
The BPT diagrams with the classifying scheme (dashed lines, from left to right):
 SF demarcations from \citet{Kewley01, Kauffmann+03};
 SF \citep{Kewley01} and AGN/LINER \citep{Kewley06} demarcations;
maximum SF value \citep{Kewley04}.
\textit{Top:} The line ratios of \INSPIRE\ points, colour-coded by the DoR, on top of predictions from SF models (grey dots). Squares are galaxies with a match in GALEX. \textit{Middle:} Emission line ratios from the shock models by \citet{Alarie09} with the parameters as described in the legend and in the text. \textit{Bottom:} Emission line ratios from the \citetalias{MartinezParedes23}  photoionization models with parameters as shown in the legend. 
}
\label{fig:bpt}
\end{figure*}

\begin{figure}
\includegraphics[width=\columnwidth]{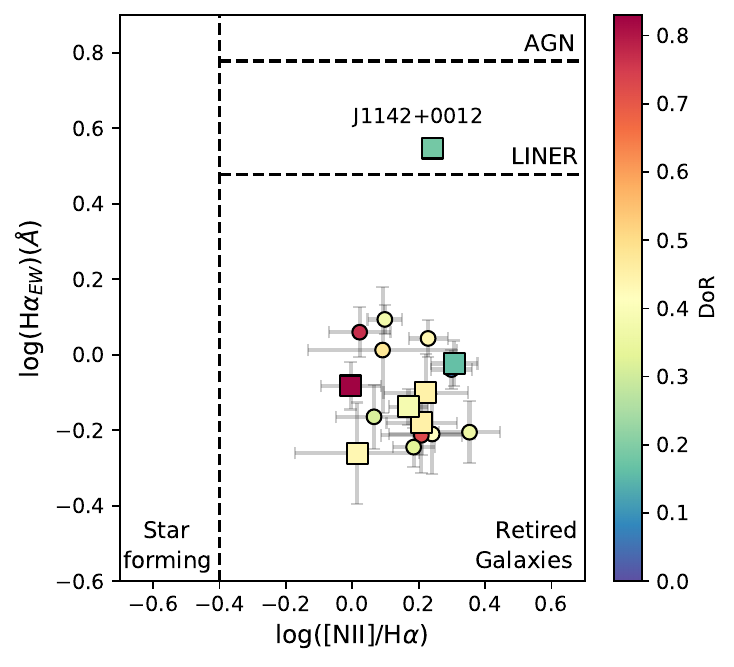}
\caption{The WHAN diagram \citep{Cid11} for the \INSPIRE\ galaxies, colour-coded by their DoR. Squares are objects with a match in GALEX. Dashed lines separate SF from LINERs and AGNs. The only system consistent with being a LINER is J1142+0012.}
\label{fig:whan}
\end{figure}

\subsection{BPT and WHAN diagnostic diagrams}
We use the classical emission-line diagnostics diagram \citep{Baldwin81, Rola97, Kauffmann+03} to further investigate on the possible source generating emission lines in red UCMGs, including in those classified as relics ($0.34<$DoR$<0.7$) and extreme relics (DoR$\ge0.7$).
This approach effectively differentiates between a star-forming (SF) and an active galactic nucleus (AGN) origin by examining the ratios of various emission lines.
We start by producing the classical 'Baldwin, Phillips \& Terlevich' (BPT) diagrams \citep{Baldwin81}.  
In these diagrams, two main tracks of galaxies are visible and separate. Star-forming objects will dominate the left-bottom region, whereas 'AGN-like' objects will lie on the right side. 
Traditionally, AGNs can then be further divided into subcategories occupying different positions in the diagrams: Seyfert galaxies in the upper part and LINERs towards the bottom region. 
Both theoretical and empirical classifications have been reported in the literature \citep[e.g.][]{Kewley01,Kewley04,Kewley06,Kauffmann+03}.
In the upper panel of Figure~\ref{fig:bpt} we show these diagrams for the \INSPIRE\ objects where emission lines are securely detected, again colour-coded by their DoR. The large uncertainty on the H$\beta$ emission, mainly due to the low SNR in the blue, prevents us from obtaining an estimate of the extinction from the Balmer decrement.
For comparison, in the same panel, we also show the line ratios
produced by SF models with an age $\le$ 3 Myr, corresponding to ionisation by OB stars (grey dots). Clearly, the \INSPIRE\ objects are not well reproduced by SF models,  having a too high [NII]/H$\alpha$ and [SII]/H$\alpha$ ratios. The SF models could still reproduce the observed [OII]/H$\alpha$ ratio (up to $\log{\mathrm{[OII]/H}\alpha}=0.32$, the maximum value allowed by the SF models,  as shown by the vertical dotted line in Figure~\ref{fig:bpt}), but only if no correction for internal dust extinction is considered. For instance, H$\alpha$/H$\beta$=4 would correspond to a correction of +0.27. 
The UCMG line ratios can instead be reproduced by shock and/or photoionisation models that include the contribution of pAGB stars, as shown in the middle and low panels of Figure~\ref{fig:bpt}. 

First, we consider the shock models from \citet{Alarie09}, available in the Mexican Million Database \citep[3MdB: ][]{Morisset2015}\footnote{\url{https://sites.google.com/site/mexicanmillionmodels/}}. The models were derived using the \textsc{MAPPINGS V} code \citep{Sutherland17}, which allows us to retrieve emission lines produced in a shocked gas, both with and without the presence of the so-called precursor\footnote{The precursor is where
the gas entering the shock is photoionised by the UV radiation emitted by the shocked gas.}.
The middle panel of Figure~\ref{fig:bpt} displays the line ratios produced by the shock models with parameters as described in \citet{Allen08}. In particular, we
consider: 
\begin{itemize}
    \item only pure shock models, since adding the precursor would produce  [OIII]/H$\beta$ ratios much higher than the value observed in \INSPIRE\ UCMGs; 
    \item models with solar ($Z=0.0183$, blue and light blue) and twice solar ($Z=0.0358$, red) metallicities; 
    \item models with shock velocities $v_{\rm sh}$ with values between 100 km s$^{-1}$ (small points) and 500 km s$^{-1}$ (big circle); 
    \item pre-shock densities of  $n_0=1$ cm$^{-3}$ (light blue) and $n_0=1$ cm$^{-3}$ (blue) for the solar metallicity models and $n_0$= 1 cm$^{-3}$ (the only available) for the super-solar models; 
    \item models with transverse magnetic fields of $10^{-4}$, 1, 5 , 10 $\mu$G cm$^{3/2}$.
\end{itemize}

Secondly, we use the models by \citet[][MP23 hereafter]{MartinezParedes23} who recently presented a large set of photoionisation models, which are publicly available in the CB\_19 table of 3MdB. 
We refer to \citetalias{MartinezParedes23} for details of how the models were computed and the description of their parameters. 
In short, the photoionisation code \textsc{Cloudy} \citep[][v. 17.03]{Ferland2017} has been used to compute emission line ratios adopting as a ionising source the population synthesis models described in \citet{Plat2019} and \citet{Sanchez2022} that include the contribution from pAGB stars (also defined as HOLMES: hot low-mass evolved stars). We select from the \citetalias{MartinezParedes23}  models the ones that give line ratios close to those measured in  the \INSPIRE\ galaxies. They are displayed in the bottom panel of Figure~\ref{fig:bpt}, where the models are chosen to have the following: 
\begin{itemize}
    \item a single stellar population (SSP) with a Kroupa IMF up to 100 solar masses (RB–SSP–Kroup–MU100) and an age of 1 Gyr as the ionising source. However we note that similar line ratios would be obtained changing the age of the models up to 10 Gyr; 
    \item metallicity of $Z=0.03$ ([M/H]=0.22), as from the stellar population analysis performed in \citetalias{Spiniello+24} we conclude that UCMGs are consistent with super-solar metallicities. We note that selecting models with solar metallicities would produce BPT ratios only slightly shifted to the SF region;
    \item an ionization parameter $\log U < -3$;
    \item a density $n_H=[10,100]$ cm$^{-3}$;
    \item values of the H$\beta$ fraction defining the thickness of the ionized cloud between 0 and 1;
    \item CNO gas abundances: C/O=[0.1, 0.44],  $\Delta$N/O=[$-0.25,0,0.25$]\footnote{Defined in \citetalias{MartinezParedes23}  as the deviation from the (N/O) to (O/H) ratio as defined in \citet{Gutkin16} }, 12+$\log$ OH=$9.14$.
\end{itemize}


With the exception of J1142+0012, models with ionisation dominated by pAGB stars
allow to reproduce well the emission line ratios observed in the \INSPIRE\ UCMGs. 
Nevertheless, as shown by \citet{Stasinska08} and further discussed by \citet{Cid11}, retired galaxies (i.e. emission-line galaxies that have stopped forming stars and are ionised by hot low-mass evolved stars) have the same location in the BPT diagram as galaxies hosting weak AGNs. 
According to \citet{Lee24}, photoionisation by pAGB stars and interstellar shocks can only be distinguished with in-depth analysis, for instance using temperature predictions. 

To try to more precisely separate
pure SF galaxies, AGN hosts (strong and weak) and passive galaxies (i.e. retired, red and dead galaxies) we use the so-called WHAN diagram \citep{Cid11} showing the [NII]/H$\alpha$ ratio against the H$\alpha$ EW.  
This is plotted in Figure~\ref{fig:whan} for the \INSPIRE\ galaxies for which we could measure H$\alpha$ and [NII] emissions with high confidence. Here, all galaxies with reliable emission lines but J1142+0012 lie in the 'retired galaxies' region, hence completely ruling out SF and disfavouring an AGN origin as the most probable source of emission lines in UCMGs and relics (with no DoR dependency). 
We therefore argue that the emission lines in \INSPIRE\ UCMGs can be explained by either the presence of pAGB stars or shocks of the gas produced by the mass loss of evolved stars, which collects at the centres of the UCMGs.  

\section{UV detection}
\label{sec:UV}
To acquire an independent line of evidence on the possible origin of emission lines, in this section we look at the UV fluxes for \INSPIRE\ objects. Unfortunately, the SNR of X-Shooter spectra of single galaxies is too low in the UV. 
Hence, to investigate whether the \INSPIRE\ UCMGs have detectable UV fluxes, 
we cross-match the \INSPIRE\ catalogue with data from the Galaxy Evolution Explorer \citep{Morrissey07_GALEX}, matching sources to GALEX photometry within $10''$. 
We caution the reader that the spatial resolution of GALEX is suboptimal when matching ultracompact galaxies, with sizes smaller than the nominal survey resolution (FWHM$\approx6
^{\prime\prime}$). However, we checked that no other source was present within this radius in the
$u$ band images from SDSS or the $g$ band images from PanStarrs. Thus, we are confident that the detected emission comes from the UCMGs themselves.

Among the 52 \INSPIRE\ objects, 20 have a match, with near-UV ($NUV$) magnitudes ranging between 21.3 and 24.1, and spanning a wide range of DoR, from nonrelics (DoR$<0.34$) with an extended SFH, to extreme relics (DoR$\ge0.7$) that have assembled all their stellar mass within the first 2 Gyr after the Big Bang (\citetalias{Spiniello+24}). In addition, 5 UCMGs have also been detected in the far-UV ($FUV$) but only one of them is a relic,  according to the \INSPIRE\ classification (J0920+0212, DoR$=0.64$). 
The final block of columns of Table~\ref{tab:emission} provides $FUV$ and $NUV$ magnitudes, as well as $NUV-r$ colours. 

\begin{figure}
    \centering      \includegraphics[width=\columnwidth]{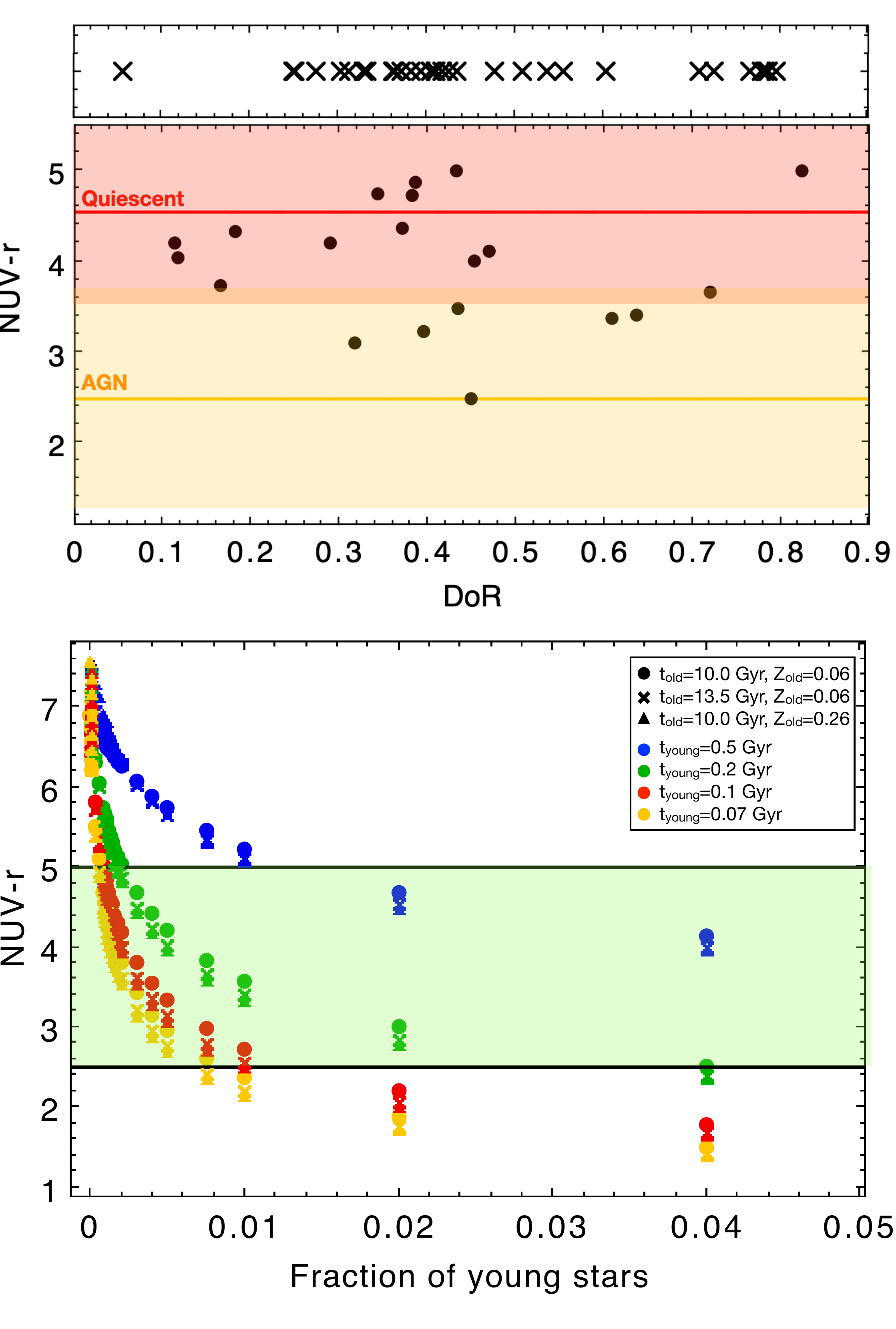}
    \caption{\textit{Top:} $NUV-r$ colour for the \INSPIRE\ galaxies with a detection in GALEX plotted versus their DoR. On the top, black crosses indicate objects without a match to show that both detection and non-detection cover a wide range of DoR. The red and yellow lines (and shaded regions) indicate the mean colour (and standard deviation) for quiescent and AGNs, respectively \citep{Ardila18}. \textit{Bottom:} $NUV-r$ prediction from E-MILES SSP \citep{Vazdekis16} old models with a small percentage of younger stars, as reported in the legend. The green shaded region shows the $NUV-r$ range covered by \INSPIRE\ galaxies.}
    \label{fig:galex}
\end{figure}

The top panel of Figure~\ref{fig:galex} shows the $NUV-r$ colour\footnote{Optical $r$-band magnitudes have been retrieved from KiDS DR4 \citep{Kuijken19_KIDSDR4}.} versus the DoR for the 20 detected objects. We also plot the objects without a match in GALEX as black crosses at the top of the panel to show that, perhaps surprisingly, no correlation with the DoR is found. 
The galaxies' colours lie perfectly in between the mean colour computed by \citet[][Tab.~1]{Ardila18} for red and dead (quiescent) galaxies ($4.54\pm1.00$, red-shaded region) and the mean colour estimated for Active Galactic Nuclei (AGNs, $2.48\pm1.24$, yellow-shaded region). 
They are consistent with the UV green valley \citep{Salim14} and lie at the redder end of those measured in post-starbust galaxies where the UV flux is mainly caused by intermediate-age pAGB stars \citep{Melnick14}. The $NUV-r$ colour of the matched \INSPIRE\ objects are instead generally redder than these computed for star-forming galaxies ($2.01\pm0.68$, \citealt{Ardila18}). 


However, another equally valid scenario can reproduce the $NUV-r$ colour of the \INSPIRE\ objects. Indeed, they can be perfectly fitted with an overall old stellar population plus the addition of a small percentage of young stars, as suggested by \citet{Salvador-Rusinol2021_nature}. This is shown in the bottom panel of Figure~\ref{fig:galex} where we plot predictions on the $NUV-r$ colour obtained from the E-MILES single stellar population models (SSPs, \citealt{Vazdekis16}).  In particular, we computed the expected colour for an old stellar population (10 Gyr) with solar metallicity to which a small percentage ($<5$ \%) of young stars ($\le 0.5$ Gyr) is added, also with solar metallicity\footnote{We stress that this assumption does not change the results as the effect of changing the metallicity on the $NUV-r$ colour is negligible.}. We explore the colour variation due to changes in age and metallicity of the main old population, finding negligible effects on the main conclusions. 

Importantly, we point out that the presence of a sub-percental population of young stars is not in disagreement with our analysis of the main ionisation mechanism of emission lines (Sec.~\ref{sec:emissionline}). In fact, the latter constrains the presence of SF on a much shorter time scale (of the order of few Myr) compared to UV colours. 

To further address this issue, we estimate the star formation rate (SFR) for the only six systems with a match in GALEX and a measured H$\alpha$ emission (J0844+0148, J0847+0112, J1218+0232, J1447-0149, J1456+0020, J2356-3332)\footnote{Excluding J1142+0012 which is an AGN.}. First, we convert the H$\alpha$ flux into a total luminosity, considering it unresolved, as the spectra are fully seeing-dominated (see \citetalias{Spiniello+24}). Second, we translate H$\alpha$ luminosities into SFRs following \citet{Kennicutt98}. In particular, we use Equation 12 and Table 1 of \citet{Kennicutt12}: 
\begin{equation}
\log\dot{\mathrm{M}}_{\star}[\mathrm{M}_{\odot} \mathrm{yr}^{-1}]= \log(\mathrm{L}_{\mathrm{H}\alpha})-\log C_x 
\end{equation}
where $\log C_x=41.27$ \citep{Murphy11}. 
The resulting SFRs for the six systems range between $0.01$ and $0.02$ $\mathrm{M}_{\odot} \mathrm{yr}^{-1}$. 

At this point, we compute the stellar mass that would be formed in $\sim$0.5 Gyr (the oldest age of the young stellar population component in Figure~\ref{fig:galex}) with a constant rate of star formation of $0.02$ $\mathrm{M}_{\odot} \mathrm{yr}^{-1}$ (as inferred from H$\alpha$), finding that it would be of the order of $\mathrm{M}_{\star,\mathrm{H}\alpha} \sim 10^{7}\mathrm{M}_{\odot}$. 
This is also in agreement with empirical relations (e.g., \citealt{Elbaz07}) and simulations (e.g., \citealt{Ciesla17}) predicting that $\mathrm{SFRs}<0.1$ correspond to $\mathrm{M}_{\star}<10^{8}\mathrm{M}_{\odot}$.  In Figure~\ref{fig:galex}, the fraction of young stars required to explain the $NUV-r$ colour depends significantly on the age of the young component, ranging from $\sim 0.001$ ($t_{young}=0.07$~Gyr) to $\sim 0.04$ ($t_{young}=0.5$~Gyr). Hence, since the typical mass of \INSPIRE\ objects is $\mathrm{M}_{\star} \ge 6\times10^{10}\mathrm{M}_{\odot}$, this implies a stellar mass in the young component $\mathrm{M}_{\star, NUV-r} \ge 6\times10^{7}\mathrm{M}_{\odot}$. 

In conclusion, since $\mathrm{M}_{\star, NUV-r}>\mathrm{M}_{\star, \mathrm{H}\alpha}$, even assuming that all the H$\alpha$ emission comes from newly formed stars ($10^{7}\mathrm{M}_{\odot}$), this is not enough to explain the $NUV-r$ colour\footnote{We note that we did not apply any correction for extinction. However, for the 5 systems for which H$\beta$ emission is also measured, the SFRs can increase up to a factor of 3, given that the Balmer decrement values range between $\sim1$ and $\sim5$ (although with very large uncertainties on H$\beta$). This is still not enough to explain the $NUV-r$ colours through the H$\alpha$ fluxes.}. 
However, we must stress a couple of important points. First, the assumption of a constant SFR might not be correct. Indeed, the SFR could decrease with time or even more likely occur in a bursty fashion (e.g., due to residual AGN activity), which would make extremely hard to detect it on very short timescales as those probed by H$\alpha$ ($\lesssim 10$ Myr). Secondly, in the case of a non-universal IMF,  within the integrated galactic stellar initial mass function (IGIMF) theory \citep{Kroupa03}, i.e. assuming that the IMF slope steepens with decreasing SFR,  the ionising photons that produce H$\alpha$ emission would be significantly lower, given the smaller number of massive stars, than for a standard IMF, meaning that the above value of $\mathrm{M}_{\star, \mathrm{H}\alpha}$ might be significantly underestimated.

\section{Discussion and Conclusions}
\label{sec:discussion}
In this paper we have reported the detection of emission lines and significant NUV emission in the UVB+VIS spectra of about half of the \INSPIRE\ UCMGs, including some of the most extreme relics. 

The [OII] emission line, which is the strongest that we detect in the \INSPIRE\ galaxies, has been widely used in the literature as an empirical SF rate indicator, calibrated through comparison with Hydrogen lines \citep[e.g.][]{Gallagher89,Rosa02,Kewley04}.
However, the situation is controversial as the same emission line, and others such as [NII] and H$\alpha$, have been detected in spectra of red and dead elliptical galaxies at different redshifts \citep{Caldwell84, Yan06, LaBarbera+19, Maseda21}, where SF is not expected. Another possible origin for emission lines in red ETGs comes from the photo-ionisation by exposed cores of evolved stars with ages greater than 100 Myr, such as pAGBs  \citep{Greggio90,Binette94, Yan12,Papaderos13, Belfiore16} and PNe \citep{Taniguchi00}. 
Alternatively, AGNs and especially LINERs, fast shock waves, and cooling flows might also produce [OII] emission \citep{Ferland83, Halpern83, Dopita95, Groves04}. 
Hence, 
we have analysed the emission line ratios to understand what is most likely causing them in the `building blocks' of massive ETGds, the oldest and densest galaxies in the nearby Universe. 



We find emission lines for galaxies at all DoR, including some of the most extreme relics. 
All but one object (J1142+0012) have a high [OII]/H$\alpha$ ratio, typical of red and dead galaxies \citepalias{Yan06}. \citet{LaBarbera+19} also found strong [OII] emission in the core of massive, giant ETGs with very old stellar populations, the region where the 'pristine' stellar population dominates the light. 
A high [OII]/H$\alpha$ is characteristic of 'retired galaxies' (\citetalias{Yan06}) and this conclusion is further corroborated by the lack of clear and strong H$\alpha$ emission, which is still present in a number of systems but very weak.  In BPT and WHAN diagrams, the line ratio of \INSPIRE\, objects cannot be reproduced by SF models.
Hence, we fully exclude that the main ionisation mechanism is on-going star formation. 
Line ratios are, instead, fully consistent with those predicted by shock models \citep{Alarie09} and by models that include the contribution from HOLMES stars \citep{MartinezParedes23}. 
Both of these scenarios advocate for internal and passive processes, rather than external or environmental ones (e.g., mergers or interactions). Importantly, this opens up new insights into the understanding of the mass assembly and cosmic evolution of the local massive ETGs.

We also looked at the $NUV-r$ colours for the 20 \INSPIRE\ galaxies with a match in GALEX. They are very similar to those measured for post-starbust galaxies \citep{Melnick14}. This may hint at the UV emission being powered by the same mechanisms (evolved stars) plus, possibly, some AGN contribution \citep{Ardila18}, which is, however, disfavoured by the line ratio analysis. It should also be pointed out that the timescales for the SF traced by emission lines and UV colours are different. In addition, the \INSPIRE\ $NUV-r$ colour can be reproduced with an overall old population ($\sim10$ Gyr) plus a sub-percent young population ($\sim 0.1$-$0.5$ Gyr). Given also the very recent results on NGC~1277 \citep{Salvador-Rusinol22} where this is indeed demonstrated from UV and VIS line-index analysis, we cannot exclude that the UV colours are due, at least partially, to a sub-percent contribution of star formation.  
In this case, since the relics did not interact with any other galaxy after the very first assembly phase at high-$z$, the emissions (and the possible residual SF causing blue UV colours) must have been caused by intrinsic processes. One possible scenario able to explain all the lines of evidence presented here is the one outlined in \citet{Dopita00}. The authors attribute the origin of the gas in the centres of red and dead massive ETGs to the mass loss due to evolved stars. In particular, pAGBs and PNe, which share the velocity dispersion of the native stellar population, give the largest contribution to the gas mass. 
A simple calculation, which assumes the PN birth rate of  ($\dot{\epsilon}\approx4\times10^{-12}$ yr$^{-1}$ L$^{-1}_{\odot}$, \citealt{Mendez97}), the typical luminosity and size of UCMGS (L$\approx6\times10^{10}$ L$_{\odot}$, \Reff$\le1.7$ kpc), and the amount of mass loss by each PN to the interstellar medium (ISM), $\sim 0.4$M$_\odot$, leads to a gas loss rate by all the PNe of about $0.24$ M$_{\odot}$ yr$^{-1}$. This means that in less than $10^{9}$ years, they can feed more than $10^{8}$ M$_{\odot}$ of gas and dust into the UCMGs. This gas, given the high relative motion of the PN envelopes through the ISM of such very dense stellar systems with high stellar velocity dispersion ($>200$ \kms, \citealt{DAgo23}), does not evaporate but collapses in dense clouds. These then infall radially towards the centre, where they can be shocked, possibly causing emission lines, and heated, therefore forming new stars at a sub-percentage level. 

The analysis we presented here allowed us to completely exclude that the emission lines are caused by star formation, but it is unfortunately unable to disentangle between shock-driven emission or pAGBs stars. Furthermore, a small contribution from the central AGN is still possible, although disfavoured. The only way forward in this sense is to  
obtain high spatial resolution spectroscopy and (far and near) UV-deep imaging to be able to resolve the stellar populations and the internal structure of the most compact and dense massive galaxies in the nearby Universe. 

\section*{Acknowledgements}
We acknowledge the usage of the Mexican Million Database \citep{Morisset2015}. 
CS, CT, FLB, DB, and PS acknowledge funding from the INAF PRIN-INAF 2020 programme 1.05.01.85.11. 
JH and CS acknowledge the financial support from the mobility programme of the Finnish Centre for Astronomy with ESO (FINCA), funded by the Academy of Finland grant nr 306531. 
AFM has received support from RYC2021-031099-I and PID2021-123313NA-I00 of MICIN/AEI/10.13039/501100011033/FEDER,UE, NextGenerationEU/PRT. CT acknowledges the INAF grant 2022 LEMON. GD acknowledges support by UKRI-STFC grants: ST/T003081/1 and ST/X001857/1.
MR acknowledges financial support from the INAF mini-grant 2022 “GALCLOCK”.

\section*{Data Availability}
The \INSPIRE\ spectra used in this paper are publicly available through the ESO Phase 3 
Archive Science Portal under the collection \INSPIRE\ (\url{https://archive.eso.org/scienceportal/home?data_collection=INSPIRE}, \url{https:https://doi.eso.org/10.18727/archive/36}). 


\bibliographystyle{mnras}
\bibliography{biblio_INSPIRE} 

\begin{thebibliography}{}
\makeatletter
\relax
\def\mn@urlcharsother{\let\do\@makeother \do\$\do\&\do\#\do\^\do\_\do\%\do\~}
\def\mn@doi{\begingroup\mn@urlcharsother \@ifnextchar [ {\mn@doi@}
  {\mn@doi@[]}}
\def\mn@doi@[#1]#2{\def\@tempa{#1}\ifx\@tempa\@empty \href
  {http://dx.doi.org/#2} {doi:#2}\else \href {http://dx.doi.org/#2} {#1}\fi
  \endgroup}
\def\mn@eprint#1#2{\mn@eprint@#1:#2::\@nil}
\def\mn@eprint@arXiv#1{\href {http://arxiv.org/abs/#1} {{\tt arXiv:#1}}}
\def\mn@eprint@dblp#1{\href {http://dblp.uni-trier.de/rec/bibtex/#1.xml}
  {dblp:#1}}
\def\mn@eprint@#1:#2:#3:#4\@nil{\def\@tempa {#1}\def\@tempb {#2}\def\@tempc
  {#3}\ifx \@tempc \@empty \let \@tempc \@tempb \let \@tempb \@tempa \fi \ifx
  \@tempb \@empty \def\@tempb {arXiv}\fi \@ifundefined
  {mn@eprint@\@tempb}{\@tempb:\@tempc}{\expandafter \expandafter \csname
  mn@eprint@\@tempb\endcsname \expandafter{\@tempc}}}

\bibitem[\protect\citeauthoryear{{Adelman-McCarthy} et~al.,}{{Adelman-McCarthy}
  et~al.}{2006}]{Adelman06}
{Adelman-McCarthy} J.~K.,  et~al., 2006, \mn@doi [\apjs] {10.1086/497917},
  \href {https://ui.adsabs.harvard.edu/abs/2006ApJS..162...38A} {162, 38}

\bibitem[\protect\citeauthoryear{{Alarie} \& {Morisset}}{{Alarie} \&
  {Morisset}}{2019}]{Alarie09}
{Alarie} A.,  {Morisset} C.,  2019, \mn@doi [\rmxaa]
  {10.22201/ia.01851101p.2019.55.02.21}, \href
  {https://ui.adsabs.harvard.edu/abs/2019RMxAA..55..377A} {55, 377}

\bibitem[\protect\citeauthoryear{{Allen}, {Groves}, {Dopita}, {Sutherland}  \&
  {Kewley}}{{Allen} et~al.}{2008}]{Allen08}
{Allen} M.~G.,  {Groves} B.~A.,  {Dopita} M.~A.,  {Sutherland} R.~S.,
  {Kewley} L.~J.,  2008, \mn@doi [\apjs] {10.1086/589652}, \href
  {https://ui.adsabs.harvard.edu/abs/2008ApJS..178...20A} {178, 20}

\bibitem[\protect\citeauthoryear{{Ardila} et~al.,}{{Ardila}
  et~al.}{2018}]{Ardila18}
{Ardila} F.,  et~al., 2018, \mn@doi [\apj] {10.3847/1538-4357/aad0a3}, \href
  {https://ui.adsabs.harvard.edu/abs/2018ApJ...863...28A} {863, 28}

\bibitem[\protect\citeauthoryear{{Baldwin}, {Phillips}  \&
  {Terlevich}}{{Baldwin} et~al.}{1981}]{Baldwin81}
{Baldwin} J.~A.,  {Phillips} M.~M.,   {Terlevich} R.,  1981, \mn@doi [\pasp]
  {10.1086/130766}, \href
  {https://ui.adsabs.harvard.edu/abs/1981PASP...93....5B} {93, 5}

\bibitem[\protect\citeauthoryear{{Barbosa}, {Spiniello}, {Arnaboldi},
  {Coccato}, {Hilker}  \& {Richtler}}{{Barbosa} et~al.}{2021}]{Barbosa21}
{Barbosa} C.~E.,  {Spiniello} C.,  {Arnaboldi} M.,  {Coccato} L.,  {Hilker} M.,
    {Richtler} T.,  2021, \mn@doi [\aap] {10.1051/0004-6361/202039809}, \href
  {https://ui.adsabs.harvard.edu/abs/2021A&A...649A..93B} {649, A93}

\bibitem[\protect\citeauthoryear{{Belfiore} et~al.,}{{Belfiore}
  et~al.}{2016}]{Belfiore16}
{Belfiore} F.,  et~al., 2016, \mn@doi [\mnras] {10.1093/mnras/stw1234}, \href
  {https://ui.adsabs.harvard.edu/abs/2016MNRAS.461.3111B} {461, 3111}

\bibitem[\protect\citeauthoryear{{Bennett}, {Larson}, {Weiland}  \&
  {Hinshaw}}{{Bennett} et~al.}{2014}]{Bennett14}
{Bennett} C.~L.,  {Larson} D.,  {Weiland} J.~L.,   {Hinshaw} G.,  2014, \mn@doi
  [\apj] {10.1088/0004-637X/794/2/135}, \href
  {https://ui.adsabs.harvard.edu/abs/2014ApJ...794..135B} {794, 135}

\bibitem[\protect\citeauthoryear{{Binette}, {Magris}, {Stasi{\'n}ska}  \&
  {Bruzual}}{{Binette} et~al.}{1994}]{Binette94}
{Binette} L.,  {Magris} C.~G.,  {Stasi{\'n}ska} G.,   {Bruzual} A.~G.,  1994,
  \aap, \href {https://ui.adsabs.harvard.edu/abs/1994A&A...292...13B} {292, 13}

\bibitem[\protect\citeauthoryear{{Buitrago}, {Trujillo}, {Conselice},
  {Bouwens}, {Dickinson}  \& {Yan}}{{Buitrago} et~al.}{2008}]{Buitrago+08}
{Buitrago} F.,  {Trujillo} I.,  {Conselice} C.~J.,  {Bouwens} R.~J.,
  {Dickinson} M.,   {Yan} H.,  2008, \mn@doi [\apjl] {10.1086/592836}, \href
  {http://adsabs.harvard.edu/abs/2008ApJ...687L..61B} {687, L61}

\bibitem[\protect\citeauthoryear{{Caldwell}}{{Caldwell}}{1984}]{Caldwell84}
{Caldwell} N.,  1984, \mn@doi [\pasp] {10.1086/131334}, \href
  {https://ui.adsabs.harvard.edu/abs/1984PASP...96..287C} {96, 287}

\bibitem[\protect\citeauthoryear{{Cid Fernandes}, {Stasi{\'n}ska}, {Mateus}  \&
  {Vale Asari}}{{Cid Fernandes} et~al.}{2011}]{Cid11}
{Cid Fernandes} R.,  {Stasi{\'n}ska} G.,  {Mateus} A.,   {Vale Asari} N.,
  2011, \mn@doi [\mnras] {10.1111/j.1365-2966.2011.18244.x}, \href
  {https://ui.adsabs.harvard.edu/abs/2011MNRAS.413.1687C} {413, 1687}

\bibitem[\protect\citeauthoryear{{Ciesla}, {Elbaz}  \& {Fensch}}{{Ciesla}
  et~al.}{2017}]{Ciesla17}
{Ciesla} L.,  {Elbaz} D.,   {Fensch} J.,  2017, \mn@doi [\aap]
  {10.1051/0004-6361/201731036}, \href
  {https://ui.adsabs.harvard.edu/abs/2017A&A...608A..41C} {608, A41}

\bibitem[\protect\citeauthoryear{{D'Ago} et~al.,}{{D'Ago}
  et~al.}{2023}]{DAgo23}
{D'Ago} G.,  et~al., 2023, \mn@doi [\aap] {10.1051/0004-6361/202245542}, \href
  {https://ui.adsabs.harvard.edu/abs/2023A&A...672A..17D} {672, A17, INSPIRE
  DR2}

\bibitem[\protect\citeauthoryear{{Daddi} et~al.,}{{Daddi}
  et~al.}{2005}]{Daddi+05}
{Daddi} E.,  et~al., 2005, \mn@doi [\apj] {10.1086/430104}, \href
  {http://adsabs.harvard.edu/abs/2005ApJ...626..680D} {626, 680}

\bibitem[\protect\citeauthoryear{{Dopita} \& {Sutherland}}{{Dopita} \&
  {Sutherland}}{1995}]{Dopita95}
{Dopita} M.~A.,  {Sutherland} R.~S.,  1995, \mn@doi [\apj] {10.1086/176596},
  \href {https://ui.adsabs.harvard.edu/abs/1995ApJ...455..468D} {455, 468}

\bibitem[\protect\citeauthoryear{{Dopita}, {Massaglia}, {Bodo}, {Arnaboldi}  \&
  {Merluzzi}}{{Dopita} et~al.}{2000}]{Dopita00}
{Dopita} M.~A.,  {Massaglia} S.,  {Bodo} G.,  {Arnaboldi} M.,   {Merluzzi} P.,
  2000, in {Kastner} J.~H.,  {Soker} N.,   {Rappaport} S.,  eds,  Astronomical
  Society of the Pacific Conference Series Vol. 199, Asymmetrical Planetary
  Nebulae II: From Origins to Microstructures. p.~423

\bibitem[\protect\citeauthoryear{{Elbaz} et~al.,}{{Elbaz}
  et~al.}{2007}]{Elbaz07}
{Elbaz} D.,  et~al., 2007, \mn@doi [\aap] {10.1051/0004-6361:20077525}, \href
  {https://ui.adsabs.harvard.edu/abs/2007A&A...468...33E} {468, 33}

\bibitem[\protect\citeauthoryear{{Ferland} \& {Netzer}}{{Ferland} \&
  {Netzer}}{1983}]{Ferland83}
{Ferland} G.~J.,  {Netzer} H.,  1983, \mn@doi [\apj] {10.1086/160577}, \href
  {https://ui.adsabs.harvard.edu/abs/1983ApJ...264..105F} {264, 105}

\bibitem[\protect\citeauthoryear{{Ferland} et~al.,}{{Ferland}
  et~al.}{2017}]{Ferland2017}
{Ferland} G.~J.,  et~al., 2017, \mn@doi [\rmxaa] {10.48550/arXiv.1705.10877},
  \href {https://ui.adsabs.harvard.edu/abs/2017RMxAA..53..385F} {53, 385}

\bibitem[\protect\citeauthoryear{{Ferr{\'e}-Mateu}, {Trujillo},
  {Mart{\'{\i}}n-Navarro}, {Vazdekis}, {Mezcua}, {Balcells}  \&
  {Dom{\'{\i}}nguez}}{{Ferr{\'e}-Mateu} et~al.}{2017}]{Ferre-Mateu+17}
{Ferr{\'e}-Mateu} A.,  {Trujillo} I.,  {Mart{\'{\i}}n-Navarro} I.,  {Vazdekis}
  A.,  {Mezcua} M.,  {Balcells} M.,   {Dom{\'{\i}}nguez} L.,  2017, \mn@doi
  [\mnras] {10.1093/mnras/stx171}, \href
  {http://adsabs.harvard.edu/abs/2017MNRAS.467.1929F} {467, 1929}

\bibitem[\protect\citeauthoryear{{Gallagher}, {Bushouse}  \&
  {Hunter}}{{Gallagher} et~al.}{1989}]{Gallagher89}
{Gallagher} J.~S.,  {Bushouse} H.,   {Hunter} D.~A.,  1989, \mn@doi [\aj]
  {10.1086/115015}, \href
  {https://ui.adsabs.harvard.edu/abs/1989AJ.....97..700G} {97, 700}

\bibitem[\protect\citeauthoryear{{Greggio} \& {Renzini}}{{Greggio} \&
  {Renzini}}{1990}]{Greggio90}
{Greggio} L.,  {Renzini} A.,  1990, \mn@doi [\apj] {10.1086/169384}, \href
  {https://ui.adsabs.harvard.edu/abs/1990ApJ...364...35G} {364, 35}

\bibitem[\protect\citeauthoryear{{Groves}, {Dopita}  \& {Sutherland}}{{Groves}
  et~al.}{2004}]{Groves04}
{Groves} B.~A.,  {Dopita} M.~A.,   {Sutherland} R.~S.,  2004, \mn@doi [\apjs]
  {10.1086/421114}, \href
  {https://ui.adsabs.harvard.edu/abs/2004ApJS..153...75G} {153, 75}

\bibitem[\protect\citeauthoryear{{Gutkin}, {Charlot}  \& {Bruzual}}{{Gutkin}
  et~al.}{2016}]{Gutkin16}
{Gutkin} J.,  {Charlot} S.,   {Bruzual} G.,  2016, \mn@doi [\mnras]
  {10.1093/mnras/stw1716}, \href
  {https://ui.adsabs.harvard.edu/abs/2016MNRAS.462.1757G} {462, 1757}

\bibitem[\protect\citeauthoryear{{Halpern} \& {Steiner}}{{Halpern} \&
  {Steiner}}{1983}]{Halpern83}
{Halpern} J.~P.,  {Steiner} J.~E.,  1983, \mn@doi [\apjl] {10.1086/184051},
  \href {https://ui.adsabs.harvard.edu/abs/1983ApJ...269L..37H} {269, L37}

\bibitem[\protect\citeauthoryear{{Kauffmann} et~al.,}{{Kauffmann}
  et~al.}{2003}]{Kauffmann+03}
{Kauffmann} G.,  et~al., 2003, \mn@doi [\mnras]
  {10.1046/j.1365-8711.2003.06292.x}, \href
  {http://adsabs.harvard.edu/abs/2003MNRAS.341...54K} {341, 54}

\bibitem[\protect\citeauthoryear{{Kennicutt}}{{Kennicutt}}{1998}]{Kennicutt98}
{Kennicutt} Jr. R.~C.,  1998, \mn@doi [\apj] {10.1086/305588}, \href
  {https://ui.adsabs.harvard.edu/abs/1998ApJ...498..541K} {498, 541}

\bibitem[\protect\citeauthoryear{{Kennicutt} \& {Evans}}{{Kennicutt} \&
  {Evans}}{2012}]{Kennicutt12}
{Kennicutt} R.~C.,  {Evans} N.~J.,  2012, \mn@doi [\araa]
  {10.1146/annurev-astro-081811-125610}, \href
  {https://ui.adsabs.harvard.edu/abs/2012ARA&A..50..531K} {50, 531}

\bibitem[\protect\citeauthoryear{{Kewley}, {Dopita}, {Sutherland}, {Heisler}
  \& {Trevena}}{{Kewley} et~al.}{2001}]{Kewley01}
{Kewley} L.~J.,  {Dopita} M.~A.,  {Sutherland} R.~S.,  {Heisler} C.~A.,
  {Trevena} J.,  2001, \mn@doi [\apj] {10.1086/321545}, \href
  {https://ui.adsabs.harvard.edu/abs/2001ApJ...556..121K} {556, 121}

\bibitem[\protect\citeauthoryear{{Kewley}, {Geller}  \& {Jansen}}{{Kewley}
  et~al.}{2004}]{Kewley04}
{Kewley} L.~J.,  {Geller} M.~J.,   {Jansen} R.~A.,  2004, \mn@doi [\aj]
  {10.1086/382723}, \href
  {https://ui.adsabs.harvard.edu/abs/2004AJ....127.2002K} {127, 2002}

\bibitem[\protect\citeauthoryear{{Kewley}, {Groves}, {Kauffmann}  \&
  {Heckman}}{{Kewley} et~al.}{2006}]{Kewley06}
{Kewley} L.~J.,  {Groves} B.,  {Kauffmann} G.,   {Heckman} T.,  2006, \mn@doi
  [\mnras] {10.1111/j.1365-2966.2006.10859.x}, \href
  {https://ui.adsabs.harvard.edu/abs/2006MNRAS.372..961K} {372, 961}

\bibitem[\protect\citeauthoryear{{Kroupa} \& {Weidner}}{{Kroupa} \&
  {Weidner}}{2003}]{Kroupa03}
{Kroupa} P.,  {Weidner} C.,  2003, \mn@doi [\apj] {10.1086/379105}, \href
  {https://ui.adsabs.harvard.edu/abs/2003ApJ...598.1076K} {598, 1076}

\bibitem[\protect\citeauthoryear{{Kuijken}}{{Kuijken}}{2011}]{Kuijken11}
{Kuijken} K.,  2011, The Messenger, \href
  {http://adsabs.harvard.edu/abs/2011Msngr.146....8K} {146, 8}

\bibitem[\protect\citeauthoryear{{Kuijken} et~al.,}{{Kuijken}
  et~al.}{2019}]{Kuijken19_KIDSDR4}
{Kuijken} K.,  et~al., 2019, \mn@doi [\aap] {10.1051/0004-6361/201834918},
  \href {https://ui.adsabs.harvard.edu/abs/2019A&A...625A...2K} {625, A2}

\bibitem[\protect\citeauthoryear{{La Barbera} et~al.,}{{La Barbera}
  et~al.}{2019}]{LaBarbera+19}
{La Barbera} F.,  et~al., 2019, \mn@doi [\mnras] {10.1093/mnras/stz2192}, \href
  {https://ui.adsabs.harvard.edu/abs/2019MNRAS.489.4090L} {489, 4090}

\bibitem[\protect\citeauthoryear{{Lee}, {Yan}, {Ji}, {Algodon}, {Westfall},
  {Lin}, {Belfiore}  \& {Bundy}}{{Lee} et~al.}{2024}]{Lee24}
{Lee} M.-Y.~L.,  {Yan} R.,  {Ji} X.,  {Algodon} G.,  {Westfall} K.,  {Lin} Z.,
  {Belfiore} F.,   {Bundy} K.,  2024, \mn@doi [\aap]
  {10.1051/0004-6361/202348459}, \href
  {https://ui.adsabs.harvard.edu/abs/2024A&A...690A..83L} {690, A83}

\bibitem[\protect\citeauthoryear{{Maksymowicz-Maciata}
  et~al.,}{{Maksymowicz-Maciata} et~al.}{2024}]{Maksymowicz-Maciata24}
{Maksymowicz-Maciata} M.,  et~al., 2024, \mn@doi [\mnras]
  {10.1093/mnras/stae1318}, \href
  {https://ui.adsabs.harvard.edu/abs/2024MNRAS.531.2864M} {531, 2864}

\bibitem[\protect\citeauthoryear{{Mart{\'\i}n-Navarro}
  et~al.,}{{Mart{\'\i}n-Navarro} et~al.}{2023}]{Martin-Navarro+23}
{Mart{\'\i}n-Navarro} I.,  et~al., 2023, \mn@doi [\mnras]
  {10.1093/mnras/stad503}, \href
  {https://ui.adsabs.harvard.edu/abs/2023MNRAS.521.1408M} {521, 1408}

\bibitem[\protect\citeauthoryear{{Mart{\'\i}nez-Paredes}, {Bruzual},
  {Morisset}, {Kim}, {Mel{\'e}ndez}  \& {Binette}}{{Mart{\'\i}nez-Paredes}
  et~al.}{2023}]{MartinezParedes23}
{Mart{\'\i}nez-Paredes} M.,  {Bruzual} G.,  {Morisset} C.,  {Kim} M.,
  {Mel{\'e}ndez} M.,   {Binette} L.,  2023, \mn@doi [\mnras]
  {10.1093/mnras/stad2447}, \href
  {https://ui.adsabs.harvard.edu/abs/2023MNRAS.525.2916M} {525, 2916}

\bibitem[\protect\citeauthoryear{{Maseda} et~al.,}{{Maseda}
  et~al.}{2021}]{Maseda21}
{Maseda} M.~V.,  et~al., 2021, \mn@doi [\apj] {10.3847/1538-4357/ac2bfe}, \href
  {https://ui.adsabs.harvard.edu/abs/2021ApJ...923...18M} {923, 18}

\bibitem[\protect\citeauthoryear{{Melnick} \& {De Propris}}{{Melnick} \& {De
  Propris}}{2013}]{Melnick14}
{Melnick} J.,  {De Propris} R.,  2013, \mn@doi [\mnras] {10.1093/mnras/stt199},
  \href {https://ui.adsabs.harvard.edu/abs/2013MNRAS.431.2034M} {431, 2034}

\bibitem[\protect\citeauthoryear{{Mendez} \& {Soffner}}{{Mendez} \&
  {Soffner}}{1997}]{Mendez97}
{Mendez} R.~H.,  {Soffner} T.,  1997, \mn@doi [\aap]
  {10.48550/arXiv.astro-ph/9611128}, \href
  {https://ui.adsabs.harvard.edu/abs/1997A&A...321..898M} {321, 898}

\bibitem[\protect\citeauthoryear{{Morisset}, {Delgado-Inglada}  \&
  {Flores-Fajardo}}{{Morisset} et~al.}{2015}]{Morisset2015}
{Morisset} C.,  {Delgado-Inglada} G.,   {Flores-Fajardo} N.,  2015, \mn@doi
  [\rmxaa] {10.48550/arXiv.1412.5349}, \href
  {https://ui.adsabs.harvard.edu/abs/2015RMxAA..51..103M} {51, 103}

\bibitem[\protect\citeauthoryear{{Morrissey} et~al.,}{{Morrissey}
  et~al.}{2007}]{Morrissey07_GALEX}
{Morrissey} P.,  et~al., 2007, \mn@doi [\apjs] {10.1086/520512}, \href
  {https://ui.adsabs.harvard.edu/abs/2007ApJS..173..682M} {173, 682}

\bibitem[\protect\citeauthoryear{{Murphy} et~al.,}{{Murphy}
  et~al.}{2011}]{Murphy11}
{Murphy} E.~J.,  et~al., 2011, \mn@doi [\apj] {10.1088/0004-637X/737/2/67},
  \href {https://ui.adsabs.harvard.edu/abs/2011ApJ...737...67M} {737, 67}

\bibitem[\protect\citeauthoryear{{Naab} et~al.,}{{Naab} et~al.}{2014}]{Naab+14}
{Naab} T.,  et~al., 2014, \mn@doi [\mnras] {10.1093/mnras/stt1919}, \href
  {https://ui.adsabs.harvard.edu/abs/2014MNRAS.444.3357N} {444, 3357}

\bibitem[\protect\citeauthoryear{{Papaderos} et~al.,}{{Papaderos}
  et~al.}{2013}]{Papaderos13}
{Papaderos} P.,  et~al., 2013, \mn@doi [\aap] {10.1051/0004-6361/201321681},
  \href {https://ui.adsabs.harvard.edu/abs/2013A&A...555L...1P} {555, L1}

\bibitem[\protect\citeauthoryear{{Plat}, {Charlot}, {Bruzual}, {Feltre},
  {Vidal-Garc{\'\i}a}, {Morisset}, {Chevallard}  \& {Todt}}{{Plat}
  et~al.}{2019}]{Plat2019}
{Plat} A.,  {Charlot} S.,  {Bruzual} G.,  {Feltre} A.,  {Vidal-Garc{\'\i}a} A.,
   {Morisset} C.,  {Chevallard} J.,   {Todt} H.,  2019, \mn@doi [\mnras]
  {10.1093/mnras/stz2616}, \href
  {https://ui.adsabs.harvard.edu/abs/2019MNRAS.490..978P} {490, 978}

\bibitem[\protect\citeauthoryear{{Rola}, {Terlevich}  \& {Terlevich}}{{Rola}
  et~al.}{1997}]{Rola97}
{Rola} C.~S.,  {Terlevich} E.,   {Terlevich} R.~J.,  1997, \mn@doi [\mnras]
  {10.1093/mnras/289.2.419}, \href
  {https://ui.adsabs.harvard.edu/abs/1997MNRAS.289..419R} {289, 419}

\bibitem[\protect\citeauthoryear{{Rosa-Gonz{\'a}lez}, {Terlevich}  \&
  {Terlevich}}{{Rosa-Gonz{\'a}lez} et~al.}{2002}]{Rosa02}
{Rosa-Gonz{\'a}lez} D.,  {Terlevich} E.,   {Terlevich} R.,  2002, \mn@doi
  [\mnras] {10.1046/j.1365-8711.2002.05285.x}, \href
  {https://ui.adsabs.harvard.edu/abs/2002MNRAS.332..283R} {332, 283}

\bibitem[\protect\citeauthoryear{{Roy} et~al.,}{{Roy} et~al.}{2018}]{Roy+18}
{Roy} N.,  et~al., 2018, \mn@doi [\mnras] {10.1093/mnras/sty1917}, \href
  {http://adsabs.harvard.edu/abs/2018MNRAS.480.1057R} {480, 1057}

\bibitem[\protect\citeauthoryear{{Salim}}{{Salim}}{2014}]{Salim14}
{Salim} S.,  2014, \mn@doi [Serbian Astronomical Journal]
  {10.2298/SAJ1489001S}, \href
  {https://ui.adsabs.harvard.edu/abs/2014SerAJ.189....1S} {189, 1}

\bibitem[\protect\citeauthoryear{{Salvador-Rusi{\~n}ol}, {Beasley}, {Vazdekis}
  \& {Barbera}}{{Salvador-Rusi{\~n}ol}
  et~al.}{2021}]{Salvador-Rusinol2021_nature}
{Salvador-Rusi{\~n}ol} N.,  {Beasley} M.~A.,  {Vazdekis} A.,   {Barbera} F.~L.,
   2021, \mn@doi [\mnras] {10.1093/mnras/staa3419}, \href
  {https://ui.adsabs.harvard.edu/abs/2021MNRAS.500.3368S} {500, 3368}

\bibitem[\protect\citeauthoryear{{Salvador-Rusi{\~n}ol}, {Ferr{\'e}-Mateu},
  {Vazdekis}  \& {Beasley}}{{Salvador-Rusi{\~n}ol}
  et~al.}{2022}]{Salvador-Rusinol22}
{Salvador-Rusi{\~n}ol} N.,  {Ferr{\'e}-Mateu} A.,  {Vazdekis} A.,   {Beasley}
  M.~A.,  2022, \mn@doi [\mnras] {10.1093/mnras/stac2070}, \href
  {https://ui.adsabs.harvard.edu/abs/2022MNRAS.515.4514S} {515, 4514}

\bibitem[\protect\citeauthoryear{{S{\'a}nchez} et~al.,}{{S{\'a}nchez}
  et~al.}{2022}]{Sanchez2022}
{S{\'a}nchez} S.~F.,  et~al., 2022, \mn@doi [\apjs] {10.3847/1538-4365/ac7b8f},
  \href {https://ui.adsabs.harvard.edu/abs/2022ApJS..262...36S} {262, 36}

\bibitem[\protect\citeauthoryear{Scognamiglio et~al.,}{Scognamiglio
  et~al.}{2020}]{Scognamiglio20}
Scognamiglio D.,  et~al., 2020, \mn@doi [The Astrophysical Journal]
  {10.3847/1538-4357/ab7db3}, 893, 4

\bibitem[\protect\citeauthoryear{{Spiniello} et~al.,}{{Spiniello}
  et~al.}{2021a}]{Spiniello20_Pilot}
{Spiniello} C.,  et~al., 2021a, \mn@doi [\aap] {10.1051/0004-6361/202038936},
  \href {https://ui.adsabs.harvard.edu/abs/2021A&A...646A..28S} {646, A28,
  INSPIRE Pilot}

\bibitem[\protect\citeauthoryear{{Spiniello} et~al.,}{{Spiniello}
  et~al.}{2021b}]{Spiniello+21}
{Spiniello} C.,  et~al., 2021b, \mn@doi [\aap] {10.1051/0004-6361/202140856},
  \href {https://ui.adsabs.harvard.edu/abs/2021A&A...654A.136S} {654, A136,
  INSPIRE DR1}

\bibitem[\protect\citeauthoryear{{Spiniello} et~al.,}{{Spiniello}
  et~al.}{2024}]{Spiniello+24}
{Spiniello} C.,  et~al., 2024, \mn@doi [\mnras] {10.1093/mnras/stad3703}, \href
  {https://ui.adsabs.harvard.edu/abs/2024MNRAS.527.8793S} {527, 8793}

\bibitem[\protect\citeauthoryear{{Stasi{\'n}ska} et~al.,}{{Stasi{\'n}ska}
  et~al.}{2008}]{Stasinska08}
{Stasi{\'n}ska} G.,  et~al., 2008, \mn@doi [\mnras]
  {10.1111/j.1745-3933.2008.00550.x}, \href
  {https://ui.adsabs.harvard.edu/abs/2008MNRAS.391L..29S} {391, L29}

\bibitem[\protect\citeauthoryear{{Sutherland} \& {Dopita}}{{Sutherland} \&
  {Dopita}}{2017}]{Sutherland17}
{Sutherland} R.~S.,  {Dopita} M.~A.,  2017, \mn@doi [\apjs]
  {10.3847/1538-4365/aa6541}, \href
  {https://ui.adsabs.harvard.edu/abs/2017ApJS..229...34S} {229, 34}

\bibitem[\protect\citeauthoryear{{Taniguchi}, {Shioya}  \&
  {Murayama}}{{Taniguchi} et~al.}{2000}]{Taniguchi00}
{Taniguchi} Y.,  {Shioya} Y.,   {Murayama} T.,  2000, \mn@doi [\aj]
  {10.1086/301520}, \href
  {https://ui.adsabs.harvard.edu/abs/2000AJ....120.1265T} {120, 1265}

\bibitem[\protect\citeauthoryear{{Tortora} et~al.,}{{Tortora}
  et~al.}{2016}]{Tortora+16_compacts_KiDS}
{Tortora} C.,  et~al., 2016, \mn@doi [\mnras] {10.1093/mnras/stw184}, \href
  {http://adsabs.harvard.edu/abs/2016MNRAS.457.2845T} {457, 2845}

\bibitem[\protect\citeauthoryear{{Tortora} et~al.,}{{Tortora}
  et~al.}{2018}]{Tortora+18_UCMGs}
{Tortora} C.,  et~al., 2018, \mn@doi [\mnras] {10.1093/mnras/sty2564}, \href
  {http://adsabs.harvard.edu/abs/2018MNRAS.481.4728T} {481, 4728}

\bibitem[\protect\citeauthoryear{{Trujillo}, {Conselice}, {Bundy}, {Cooper},
  {Eisenhardt}  \& {Ellis}}{{Trujillo} et~al.}{2007}]{Trujillo+07}
{Trujillo} I.,  {Conselice} C.~J.,  {Bundy} K.,  {Cooper} M.~C.,  {Eisenhardt}
  P.,   {Ellis} R.~S.,  2007, \mn@doi [\mnras]
  {10.1111/j.1365-2966.2007.12388.x}, \href
  {http://adsabs.harvard.edu/abs/2007MNRAS.382..109T} {382, 109}

\bibitem[\protect\citeauthoryear{{Trujillo}, {Cenarro}, {de
  Lorenzo-C{\'a}ceres}, {Vazdekis}, {de la Rosa}  \& {Cava}}{{Trujillo}
  et~al.}{2009}]{Trujillo+09_superdense}
{Trujillo} I.,  {Cenarro} A.~J.,  {de Lorenzo-C{\'a}ceres} A.,  {Vazdekis} A.,
  {de la Rosa} I.~G.,   {Cava} A.,  2009, \mn@doi [\apjl]
  {10.1088/0004-637X/692/2/L118}, \href
  {http://adsabs.harvard.edu/abs/2009ApJ...692L.118T} {692, L118}

\bibitem[\protect\citeauthoryear{{Trujillo}, {Ferr{\'e}-Mateu}, {Balcells},
  {Vazdekis}  \& {S{\'a}nchez-Bl{\'a}zquez}}{{Trujillo}
  et~al.}{2014}]{Trujillo14}
{Trujillo} I.,  {Ferr{\'e}-Mateu} A.,  {Balcells} M.,  {Vazdekis} A.,
  {S{\'a}nchez-Bl{\'a}zquez} P.,  2014, \mn@doi [\apjl]
  {10.1088/2041-8205/780/2/L20}, \href
  {http://adsabs.harvard.edu/abs/2014ApJ...780L..20T} {780, L20}

\bibitem[\protect\citeauthoryear{{Vazdekis}, {Koleva}, {Ricciardelli},
  {R{\"o}ck}  \& {Falc{\'o}n-Barroso}}{{Vazdekis} et~al.}{2016}]{Vazdekis16}
{Vazdekis} A.,  {Koleva} M.,  {Ricciardelli} E.,  {R{\"o}ck} B.,
  {Falc{\'o}n-Barroso} J.,  2016, \mn@doi [\mnras] {10.1093/mnras/stw2231},
  \href {https://ui.adsabs.harvard.edu/abs/2016MNRAS.463.3409V} {463, 3409}

\bibitem[\protect\citeauthoryear{{Vernet} et~al.,}{{Vernet}
  et~al.}{2011}]{Vernet11}
{Vernet} J.,  et~al., 2011, \mn@doi [\aap] {10.1051/0004-6361/201117752}, \href
  {https://ui.adsabs.harvard.edu/abs/2011A&A...536A.105V} {536, A105}

\bibitem[\protect\citeauthoryear{{Yan} \& {Blanton}}{{Yan} \&
  {Blanton}}{2012}]{Yan12}
{Yan} R.,  {Blanton} M.~R.,  2012, \mn@doi [\apj] {10.1088/0004-637X/747/1/61},
  \href {https://ui.adsabs.harvard.edu/abs/2012ApJ...747...61Y} {747, 61}

\bibitem[\protect\citeauthoryear{{Yan}, {Newman}, {Faber}, {Konidaris}, {Koo}
  \& {Davis}}{{Yan} et~al.}{2006}]{Yan06}
{Yan} R.,  {Newman} J.~A.,  {Faber} S.~M.,  {Konidaris} N.,  {Koo} D.,
  {Davis} M.,  2006, \mn@doi [\apj] {10.1086/505629}, \href
  {https://ui.adsabs.harvard.edu/abs/2006ApJ...648..281Y} {648, 281}

\bibitem[\protect\citeauthoryear{{de Jong} et~al.,}{{de Jong}
  et~al.}{2017}]{deJong+17_KiDS_DR3}
{de Jong} J.~T.~A.,  et~al., 2017, \mn@doi [\aap]
  {10.1051/0004-6361/201730747}, \href
  {http://adsabs.harvard.edu/abs/2017A%26A...604A.134D} {604, A134}

\bibitem[\protect\citeauthoryear{{van Dokkum} et~al.,}{{van Dokkum}
  et~al.}{2008}]{vanDokkum+08}
{van Dokkum} P.~G.,  et~al., 2008, \mn@doi [\apjl] {10.1086/587874}, \href
  {http://adsabs.harvard.edu/abs/2008ApJ...677L...5V} {677, L5}

\makeatother
\end{thebibliography}


\bsp	
\label{lastpage}

\end{document}